\documentclass[%
 reprint,
 amsmath,amssymb,
 aps,
]{revtex4-2}

\usepackage{graphicx}% Include figure files
\usepackage{dcolumn}% Align table columns on decimal point
\usepackage{bm}% bold math
\usepackage[detect-all]{siunitx}
\DeclareSIUnit\gauss{G}
\usepackage{svg}
\usepackage{soul}
\usepackage{ulem}
\usepackage{xr}

\newcommand{\mfa}[1]{\textcolor{black}{#1}}

\usepackage{sidecap}

\begin{document}

\preprint{APS/123-QED}

\title{\mfa{Spin-entanglement of an atomic pair through coupling to their thermal motion}}

\author{Poramaporn Ruksasakchai}
\affiliation{Department of Physics, University of Otago, Dunedin 9016, New Zealand}
\affiliation{The Dodd-Walls Centre for Photonic and Quantum Technologies, University of Otago, Dunedin 9016, New Zealand}

\author{Lucile Sanchez}
\affiliation{Department of Physics, University of Otago, Dunedin 9016, New Zealand}
\affiliation{The Dodd-Walls Centre for Photonic and Quantum Technologies, University of Otago, Dunedin 9016, New Zealand}

\author{Marvin Weyland}
\affiliation{Department of Physics, University of Otago, Dunedin 9016, New Zealand}
\affiliation{The Dodd-Walls Centre for Photonic and Quantum Technologies, University of Otago, Dunedin 9016, New Zealand}

\author{Scott Parkins}
\affiliation{Department of Physics, University of Auckland, Auckland 1010, New Zealand}
\affiliation{The Dodd-Walls Centre for Photonic and Quantum Technologies, University of Otago, Dunedin 9016, New Zealand}

\author{Stuart S. Szigeti}
\affiliation{Department of Quantum Science and Technology, Research School of Physics, The Australian National University, Canberra 2601, Australia}

\author{Mikkel F. Andersen}
\affiliation{Department of Physics, University of Otago, Dunedin 9016, New Zealand}
\affiliation{The Dodd-Walls Centre for Photonic and Quantum Technologies, University of Otago, Dunedin 9016, New Zealand}

\date{\today}% It is always \today, today,

\begin{abstract}
The spin-dynamics of two alkali atoms in an optical tweezer is driven by spin-changing collisions that couple the spin-state of the atoms to their relative motion. \mfa{This paper} experimentally \mfa{studies the resulting spin-states} %demonstrate that, 
when the relative motion is in a thermal state \mfa{with $k_B T$ much larger than the energies of the spin-states that take part in the dynamics. We find that }%, this coupling can drive 
an initially unentangled spin-state \mfa{can evolve} into an entangled state.
\mfa{This is contrary to the common case when coupling a quantum system to hot degrees of freedom leads to loss of entanglement and not its generation.}
%This is contrary to the common case where coupling a quantum system to thermal degrees of freedom leads to loss of entanglement.  
Moreover, we show that the generated entanglement is technologically useful as it, in principle, can enhance the sensitivity of measurements beyond the standard quantum limit. 
This may provide a promising avenue for robust entanglement generation for future technologies.
\end{abstract}

\maketitle

Quantum entanglement could enable future revolutionary technologies for faster computing~\cite{Ladd:2010, Preskill:2018, Leon:2021}, more secure communications~\cite{Pirandola:2020,Sidhu:2021}, and more sensitive measurements~\cite{Pezze:2018, Szigeti:2021}.
However, the fragility of entanglement has impeded the development and widespread uptake of entanglement-enhanced technologies. In particular, any interaction of a quantum system with a thermal environment is generally considered detrimental to entanglement-enhanced technologies, due to the rapid loss of entanglement caused by this interaction.  
To date, exploiting entanglement has therefore required the careful shielding of quantum systems from thermal environments -- generally an arduous task -- or protecting the state in so-called decoherence-free sub-spaces~\cite{Kwiat2000}. Yet, coupling to an environment also drives evolution in a quantum system, and under certain conditions such dissipative dynamics can cause the quantum system to evolve into an entangled state~\cite{Braun2002, Kim2002, Parkins2006, Pielawa2007, Diehl2008, Kastoryano2011, Foss2012, Carr2013, Reiter2013, Chen2017, He2019, Liu2021, Wang2023}. 
Experimental demonstrations that generate useful entanglement through dissipative dynamics have focused on situations when the environment is in a pure state 
\cite{Krauter2011, Barreiro2011, Lin2013, Shankar2013, Passos2018, Wang2020}. 
Explorations into whether technologically-useful entanglement can be generated through thermalization dynamics, which normally destroy entanglement, 
are therefore lacking.   

Here, we experimentally demonstrate that a coupling to thermal degrees of freedom can dissipatively drive a quantum system from an unentangled state into an entangled one. Furthermore, we demonstrate that the entanglement generated is technologically useful and could enable sensitive magnetic field measurements beyond the standard quantum limit. Our experiments exploit conservation rules that prohibit complete thermalization and ensure only entangled states are populated. To achieve this, we utilize the extraordinary level of flexibility and control provided by the manipulation of individual atoms using optical tweezers~\cite{Andersen2022}. Explicitly, we study the entanglement in two-body spin-states that emerges as a result of dissipative spin-dynamics driven by the coupling between the spin and motional degrees of freedom of two atoms in an optical tweezer. The dominant energy scale in the two-atom system is $k_B T$ where $k_B$ is Boltzmann's constant, and $T$ is the temperature associated with the atoms' motional degrees of freedom. The atomic motion is usually considered classical in this regime, since $k_B T$ is significantly larger than both the energy-level separation associated with the atoms' motion in the trap and the energy differences between atomic spin-states partaking in the dynamics. Nevertheless, our experiments demonstrate that entanglement can be robustly generated even in such a traditional classical regime. 

\textit{Entanglement via spin-changing collisions.---}We consider the spin states of two thermal $^{85}$Rb atoms in an optical tweezer initially in their $F=2,\,m_F=0$ groundstates (denoted $\left|0\right\rangle$ from here on). \mfa{A harmonic approximation of the tweezer potential allows for separation of center of mass and relative coordinates, which decouples the center of mass dynamics from the spin-dynamics. The Hamiltonian governing the spin-dynamics is \cite{sompet2019}}
\begin{equation}
    \hat{H}=\frac{\hat{\mathbf{p}}^2}{2 \mu} + \sum_{j=x,y,z}{\frac{1}{2} \mu \omega_j \hat{r}_j^2}+ \sum_{i=1,2}{\hat{H}_{Z,i}} + \hat{H}_S,
\end{equation}
\mfa{where $\hat{\mathbf{r}}$ is the relative position, $\hat{\mathbf{p}}$ the relative momentum, $\mu$ the reduced mass, $\omega_j$ the oscillation frequencies in the tweezer, $\hat{H}_{Z,i}$ the Zeeman energy of atom $i$, and $\hat{H}_S$ the atom-atom interaction Hamiltonian. In the absence of $\hat{H}_S$ the system separates into relative motion ($\hat{H}_r=\frac{\hat{\mathbf{p}}^2}{2 \mu} + \sum_{j=x,y,z}{\frac{1}{2} \mu \omega_j \hat{r}_j^2}$) and spin $\hat{H}_Z=\sum_{i=1,2}{\hat{H}_{Z,i}}$}.  
The atoms' \mfa{relative} motional states provide thermal degrees of freedom that couple to the atoms' spin through \mfa{$\hat{H}_S$, which gives} rise to spin-changing collisions. \mfa{The coupling Hamiltonian is a product of a relative motion operator and a spin-operator and} has the form~\cite{Ho1998, sompet2019}:
\begin{equation}
    \hat{H}_S=V\left( \hat{\mathbf{r}} \right)\times \sum_{m_1,m_2,m'_1,m'_2}{g_{m_1, m_2}^{m'_1, m'_2}\left|m'_1, m'_2 \right\rangle \left\langle m_1, m_2 \right|}, \label{hs}
\end{equation}
where %$\hat{\mathbf{r}}$ is the relative position operator of the two atoms, 
$V\left( \hat{\mathbf{r}} \right)$ is the atom-atom interaction potential, and $\left|m_1, m_2 \right\rangle$ a spin-state where atom 1 (2) has a spin projection on the $z$-axis given by $m_1$ ($m_2$). $g_{m_1, m_2}^{m'_1, m'_2}$ characterizes the coupling strength from $\left|m_1, m_2 \right\rangle$ to $\left|m'_1, m'_2 \right\rangle$. Importantly, $g_{m_1, m_2}^{m'_1, m'_2}=0$ if $m_1+m_2 \neq m'_1+m'_2$, so $\left|0, 0 \right\rangle$ is only coupled to states with $m'_1+m'_2=0$ \cite{Ho1998}. The interaction in Eq.~(\ref{hs}) can generate entanglement in ultra-cold systems through coherent evolution when atoms are in a pure motional state \cite{gross2011, luo2017, fadel2018, kunkel2018, lange2018, kaufman2015}. However, it can also drive interesting spin dynamics when the atomic motion is thermally excited \cite{Pechkis2013}, and these dynamics lead to strong correlations between the spin-states of the atoms \cite{sompet2019}. 

Two conservation rules restrict the relaxation dynamics to a subspace of spin states consisting of the initial unentangled state ($\left| 0,0 \right\rangle$) and two entangled spin-states. % that are populated through the dynamics. 
The first conservation rule is the above mentioned $m_1+m_2=0$, restricting accessible spin-states to those with $m_1=-m_2$. The second conservation rule arises due to the spherical symmetry of the atom-atom interaction potential $V\left( \hat{\mathbf{r}} \right)$ in Eq.~(\ref{hs}), which conserves the parity of the relative wavefunction. The relative wavefunction must have even parity since the atoms are bosons and initially in the same internal state. 
Since the parity is conserved, the spin wavefunction must remain symmetric under exchange. The relaxation dynamics driven by the coupling of the spin-states to the thermal relative motion can therefore only populate the entangled spin-states $\left| \chi_1 \right\rangle=\frac{1}{\sqrt{2}}\left( \left|1, -1 \right\rangle + \left|-1, 1 \right\rangle \right)$ and $\left| \chi_2 \right\rangle=\frac{1}{\sqrt{2}}\left( \left|2, -2 \right\rangle + \left|-2, 2 \right\rangle \right)$, as these are the only spin-states that fulfill the conservation rules. \mfa{ Note that, since it is conservation rules that restrict the relaxation dynamics to a subspace of entangled spin-states, it is unaffected by the relative motion temperature. We therefore conduct our investigation in the regime when $k_BT$ is much larger than the energies of the spin-states involved in the dynamics. In this regime, there is no significant coherence between $\left| 0,0 \right\rangle$, $\left| \chi_1 \right\rangle$, and $\left| \chi_2 \right\rangle$, yet $\left| \chi_1 \right\rangle$ and $\left| \chi_2 \right\rangle$ are entangled states. If the atomic pair had been prepared in the motional ground state, the coupling of the spins to this particular state would still drive dynamics, but in this case we would expect all aspects of the dynamics to be coherent. In either case, the individual atoms' spins do not get entangled with the thermal degrees of freedom. This is critical, as such entanglement would generally lead to a mixed two-body spin-state.   
Finally,}     
%Crucially, 
$\left| \chi_1 \right\rangle$ and $\left| \chi_2 \right\rangle$ are insensitive to magnetic noise, which generally causes rapid dephasing of atomic spin states. The entangled states are therefore born in a decoherence-free subspace, preserving them for later use.

\begin{figure}
  \includegraphics[scale=0.5]{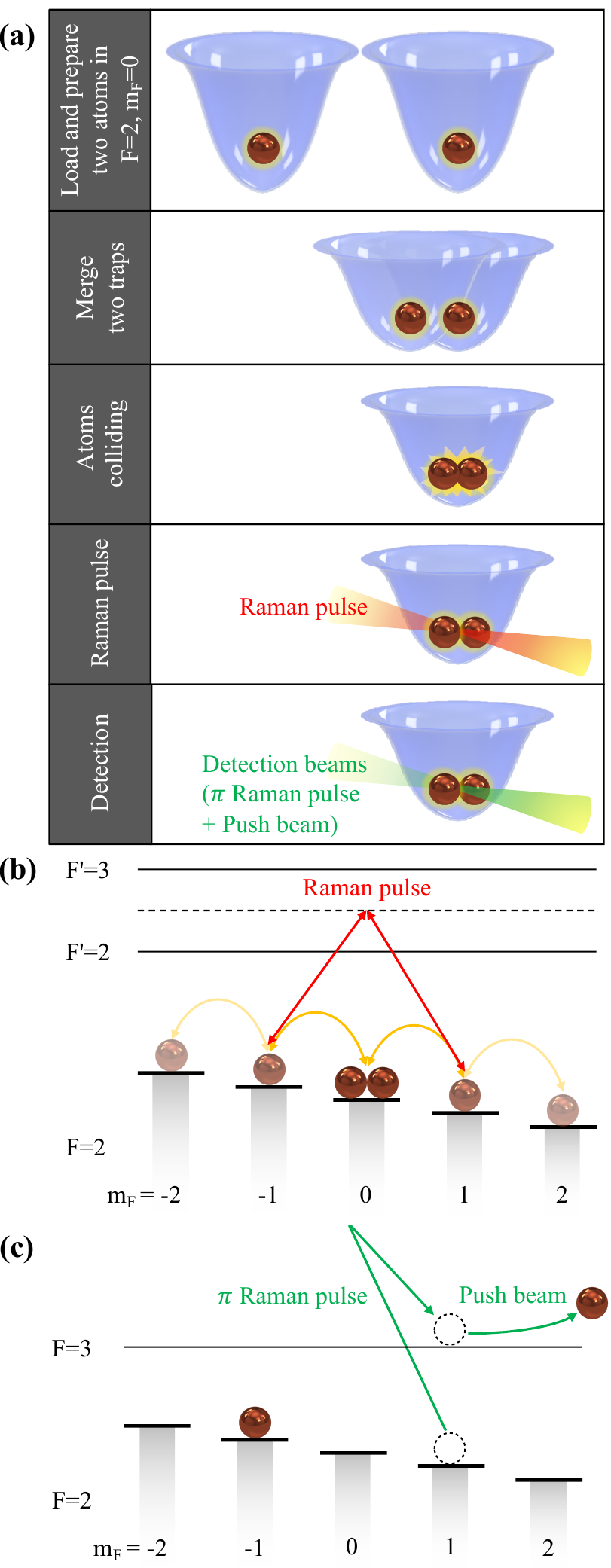}
	\caption{Experimental procedure for entanglement generation and investigation. \textbf{a}: Two atoms are initially loaded and prepared in the $F=2, m_F=0$ state in separate optical tweezers. By merging the tweezers, both atoms end up in the same tweezer. This allows the spin-changing collisions illustrated in \textbf{b}. A Raman pulse (also illustrated in \textbf{b}) couples $|-1\rangle$ and $|1\rangle$. Finally, atoms in $|1\rangle$ are ejected (see \textbf{c}) and the remaining atom number distribution is determined.
    \textbf{b}: Two atoms initially in $|0,0\rangle$ couple to the states $|\chi_1\rangle$ (dark yellow arrows) and $|\chi_2\rangle$ (light yellow arrows) by spin-changing collisions.  Red arrows illustrate the Raman pulse. \textbf{c}: The atoms in $|1\rangle$ are ejected using a Raman pulse and push beam.  
 }
        \label{fig:Process}
\end{figure}

\textit{Experiment.---}The experimental procedure is outlined in Fig.~\ref{fig:Process} (a), \mfa{and its technical details are in the appendix. We initially load two separate tweezers, with one atom each \cite{carpentier2013,fung2015, grunzweig2010}, and prepare the atoms in $\left| 0 \right\rangle$. Merging the tweezers leads to both atoms residing in a $h\times$\SI{22}{\mega\hertz} deep tweezer, with a temperature of $\sim$\SI{40}{\micro\kelvin}. At this stage the bias magnetic field is \SI{0.5}{\gauss}, and the atom-atom interaction generates the $m_F$ population dynamics illustrated in Fig.~\ref{fig:Process} (b).}
After \SI{100}{\milli\second}, 
the atoms show an appreciable probability of occupying the $|1\rangle$ and $|-1\rangle$ states (see Fig.~S1~\cite{supplementary}). 
After ramping the bias magnetic field to \SI{8.5}{\gauss}, a Raman pulse of duration $\tau$ couples the $|1\rangle$ and $|-1\rangle$ states (the red arrows in Fig.~\ref{fig:Process} (b)). \mfa{To determine the resulting spin-state, we then expel atoms in state $|1\rangle$ from the trap and determine the number remaining ~\cite{hilliard2015}, which must have been in $|-1\rangle$ when an atom is expelled.}

\textit{Entanglement generation and validation.---} \mfa{Previous work has demonstrated highly correlated spin-states as a result of spin-changing collisions in pairs of atoms in an optical tweezer \cite{sompet2019}. However, strong correlations are insufficient to prove entanglement, since unentangled states like $\left|1, -1 \right\rangle$ and $\left|-1, 1 \right\rangle$, or any classical mixture of the two, also produce perfect correlations. Proof of entanglement requires a demonstration of coherence between $\left|1, -1 \right\rangle$ and $\left|-1, 1 \right\rangle$ in $\left| \chi_1 \right\rangle$.}

\begin{figure}
\centering
        \centering
        \includegraphics[scale=.52]{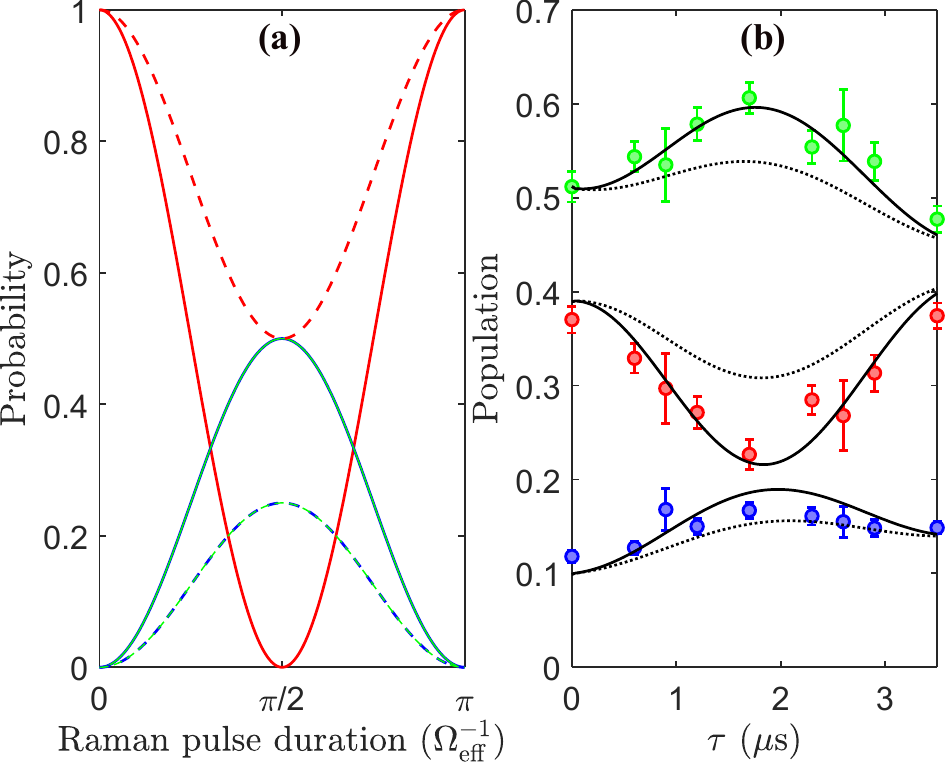}
        \caption{
        \mfa{\textbf{a}: Idealized calculation of probabilities for finding 1 (red), or 2 (green) atoms in the $F=2$, $m_F=-1$ state as a function of Raman-pulse duration. The probability for finding 0 is identical to that of finding 2. Solid and dashed curves represent entangled and unentangled states, respectively. \textbf{b}: Experimental populations of 0-atom (blue), 1-atom (red), and 2-atom (green) as a function of Raman-pulse duration $\tau$ from 0 to $\pi$ pulse. Solid and dashed curves show the corresponding theoretical expectations for entangled and unentangled states.}}        
        \label{fig:Entanglement_data}
\end{figure}

\mfa{The Raman-pulse that drives Rabi-flopping dynamics between $\left|1 \right\rangle$ and  $\left|-1 \right\rangle$ tests for coherence between $\left|1, -1 \right\rangle$ and  $\left|-1, 1 \right\rangle$. The timescale of this dynamics ($\mu$s) is much faster than the collisional dynamics timescale (ms), so we can neglect interactions during the Raman pulse, and the two atoms undergo Rabi flopping independently of each other. 
For a Raman pulse duration $\Omega t$, the individual states transform as $|\pm 1\rangle\rightarrow \cos (\Omega t/2)|\pm 1\rangle +i\sin (\Omega t/2)|\mp 1\rangle$. In general, this couples the two-atom states $|1,-1\rangle$ and $|-1,1\rangle$ to each other and to $|1,1\rangle$ and $|-1,-1\rangle$, but for $\Omega t=\pi/2$ (that is, a $\pi/2$-pulse), the entangled state $|\chi_1\rangle$ transforms as $|\chi_1\rangle\rightarrow \frac{i}{\sqrt{2}}\left( \left|1, 1 \right\rangle + \left|-1, -1 \right\rangle \right)\equiv i|\phi_+\rangle$, which results from perfect destructive interference between amplitudes for the states $|\pm 1,\mp 1\rangle$ in the superposition. Hence, the probability of measuring the atoms in different states is zero.
In contrast, one can show that the perfectly correlated but unentangled mixed state $\hat\rho_\textrm{mix}=\frac{1}{2}\left( |1,-1\rangle\langle 1,-1|+|-1,1\rangle\langle -1,1|\right)$ transforms under a $\pi/2$-pulse as $\hat\rho_\textrm{mix}\rightarrow \frac{1}{2} \left( |\phi_+\rangle\langle\phi_+| + |\chi_-\rangle\langle \chi_-|\right)$, where $|\chi_-\rangle =\frac{1}{\sqrt{2}}\left( \left|1,-1 \right\rangle - \left|-1, 1 \right\rangle \right)$, giving a 50\% probability of measuring the atoms in different states. 
As illustrated in Fig.~\ref{fig:Entanglement_data}(a), this means that as a function of the Raman pulse duration, the oscillations in the probabilities to measure atoms in the same or different states are reduced in amplitude for the initial mixed state $\hat\rho_\textrm{mix}$ by 50\% compared to the entangled state $|\chi_1\rangle$. This difference in the oscillation amplitude provides a clear means of discriminating $|\chi_1\rangle$ from an unentangled state.
}

There are several factors that prohibit the direct comparison of the experimental data with the idealized outcomes \mfa{ shown in Fig. \ref{fig:Entanglement_data}a. Most prominently is the fact that only a proportion of the atom pairs have undergone spin-changing collisions into $\left| \chi_1 \right\rangle$ with the remaining predominantly in $\left| 0,0 \right\rangle$ or $\left| \chi_2 \right\rangle$. This offsets the curves. Additional factors are:}  
imperfections in the $m_F$ state preparation and $m_F$ state detection, the finite efficiency in ejecting atoms in the $F=3$ state, the probability of atom loss, spontaneous emission during the $|1\rangle \to |-1\rangle$ Raman pulse, as well as a two-atom loss due to inelastic collisions when one of the atoms is in the $F=3$ groundstate. \mfa{For more details see~\cite{supplementary}. To compare experimental data to theoretical outcomes, we model how these factors affect the idealized results of Fig. \ref{fig:Entanglement_data}a and plot the result together with the experimental data in Fig. \ref{fig:Entanglement_data}b.}  
\mfa{The figure} displays the measured population of 0-, 1-, and 2-atoms remaining in the trap as a function of $\tau$. The solid lines represent the model with maximum entanglement \mfa{(the solid curves in Fig. \ref{fig:Entanglement_data}a)}, while the dashed lines correspond to the theoretical model with a totally mixed state \mfa{ (the dashed curves in Fig. \ref{fig:Entanglement_data}a)}. \mfa{The smaller variations in populations for the mixed state is clearly visible.} Notably, we observe that the 1-atom and 2-atom data agree with the curves that assume a fully entangled spin state (see Table ~S1~\cite{supplementary}).%; this is confirmed through chi-square tests that reject the unentangled hypothesis and accept the fully-entangled hypothesis (see Table [S1]~\cite{supplementary}).

The experimental probability that no atoms survive does not follow the model in Fig.~\ref{fig:Entanglement_data}b. A reason for this is the rudimentary way in which the simulation accounts for the two-atom loss due to inelastic collision. This process strongly contributes to the probability that both atoms are lost and is applied as a fixed probability in the simulation when one atom is in F=3 and the other is in F=2 before push out. However, inelastic collision rates also depend on the combination of $m_F$-states of the atoms, and these depend on $\tau$. We are unable to independently measure the inelastic loss probabilities for all combinations of $m_F$-states that emerge in the simulation and therefore refrain from making any conclusions based on comparison of the zero-atom curve. We note that it is possible to choose inelastic collision rates such that all curves agree with the fully entangled model, but it is not possible to get agreement with the unentangled model~\cite{supplementary}. \mfa{The 2-atom curve (green dots) is unaffected by the inelastic collisions, and we can therefore use it to quantify the fidelity of the entangled state without potentially introducing systematic errors. To do so we run the model for various degrees of entanglement and compute the $\chi^2$ value for each (see Fig.~S4~\cite{supplementary}). From this analysis we see that the best model is the fully entangled (a fidelity of 1 if the atoms have undergone spin-changing collision into $m=\pm 1$ states), and that the fidelity of the entangled state is above 0.9 with 68\% confidence. No entanglement is firmly rejected.}

\textit{Entanglement destruction.---} To further investigate the presence of entanglement, we compare to an experiment in which we deliberately attempt to destroy the generated entanglement via a two-step process. 
First, we ramp the bias magnetic field up over duration \SI{0.8}{\milli\second}, which gives rise to a small magnetic field gradient of about \SI{2.8}{\gauss / \cm}.
Second, we modulate the tweezer potential by abruptly turning on a second tweezer partially overlapped with the tweezer that holds the atoms and immediately ramping it off again over \SI{10}{\milli\second}. 
The magnetic field gradient alone is insufficient to cause a significant loss of entanglement, but it breaks the symmetry of the relative wavefunction such that it is no longer a pure even-parity state. With parity conservation broken, random mixing of states in the second step leads to occupation of relative wavefunctions that have predominantly odd parity. Odd-parity states have a spin-wavefunction of $\left| \chi_- \right\rangle=\frac{1}{\sqrt{2}}\left( \left|1, -1 \right\rangle - \left|-1, 1 \right\rangle \right)$, and an equal mixture of $\left| \chi_- \right\rangle$ and $\left| \chi_1 \right\rangle$ means that the spin-entanglement is lost.
\begin{figure} 
\centering
        \centering
        \includegraphics[scale=.52]{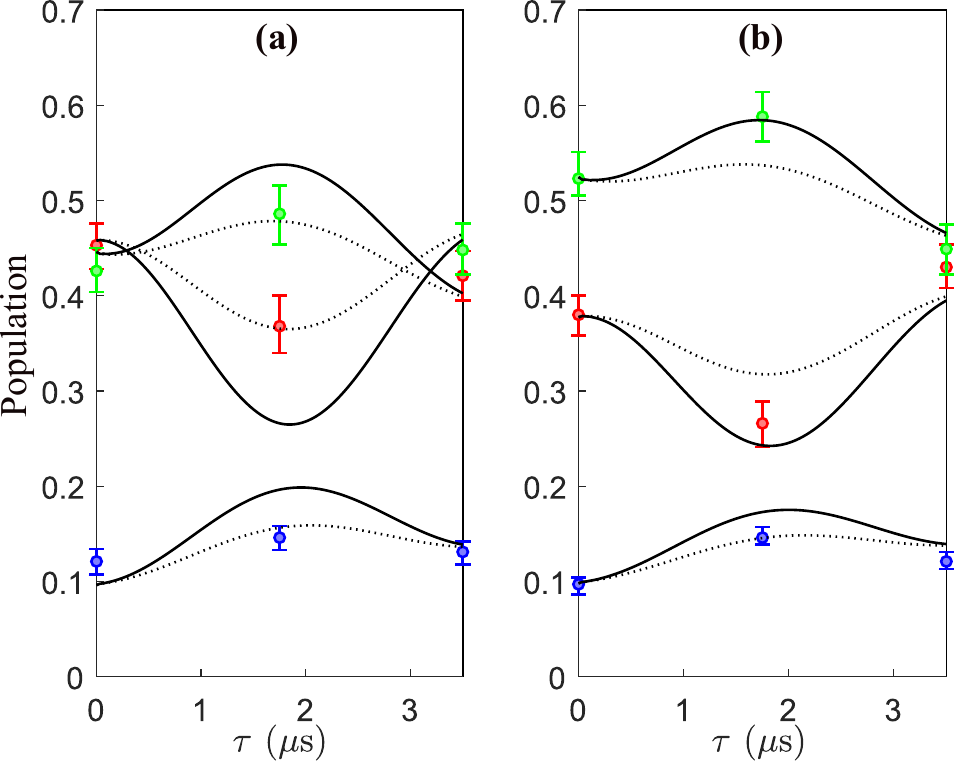}
        \caption{\textbf{a}: Data taken with the entanglement destruction procedure, \mfa{showing agreement with the unentangled model}. \textbf{b}: The same experimental procedure as in Fig. \ref{fig:Entanglement_data}b, but with the data taken interleaved with that of \textbf{a}. Symbols and curves are the same as in Fig. \ref{fig:Entanglement_data}b.}
        \label{fig:Shake}
\end{figure}

Figure~\ref{fig:Shake}a shows measurements similar to Fig.~\ref{fig:Entanglement_data}b, but with the entanglement destruction procedure applied between collisions and the Raman pulse. 
The 1- and 2-atom probabilities now agree with the unentangled model. To eliminate the possibility that drift or change in the experiment contributes to the different experimental outcome, we also acquired data without the entanglement destruction procedure. It was taken in random order and interleaved with the measurements in Fig.~\ref{fig:Shake}a, and Fig.~\ref{fig:Shake}b shows the result. It is consistent with Fig.~\ref{fig:Entanglement_data}b as expected. The agreement with the unentangled model in Fig.~\ref{fig:Shake}a is hence due to the entanglement destruction procedure. We therefore conclude that it causes loss of entanglement and that the pair is entangled prior to it. The initial populations differ with and without the entanglement destruction procedure, since heating causes some atom loss, and there is also spin-changing collision dynamics during the procedure.     

\textit{Potential for entanglement-enhanced magnetometry.---}We seek to investigate whether the spin-entanglement is metrologically useful. We do that by replacing the $|1\rangle \to |-1\rangle$ Raman pulse with a Ramsey-type $\pi/2$-wait-$\pi/2$ pulse sequence.
\begin{figure} 
\centering 
\includegraphics[scale=0.35]{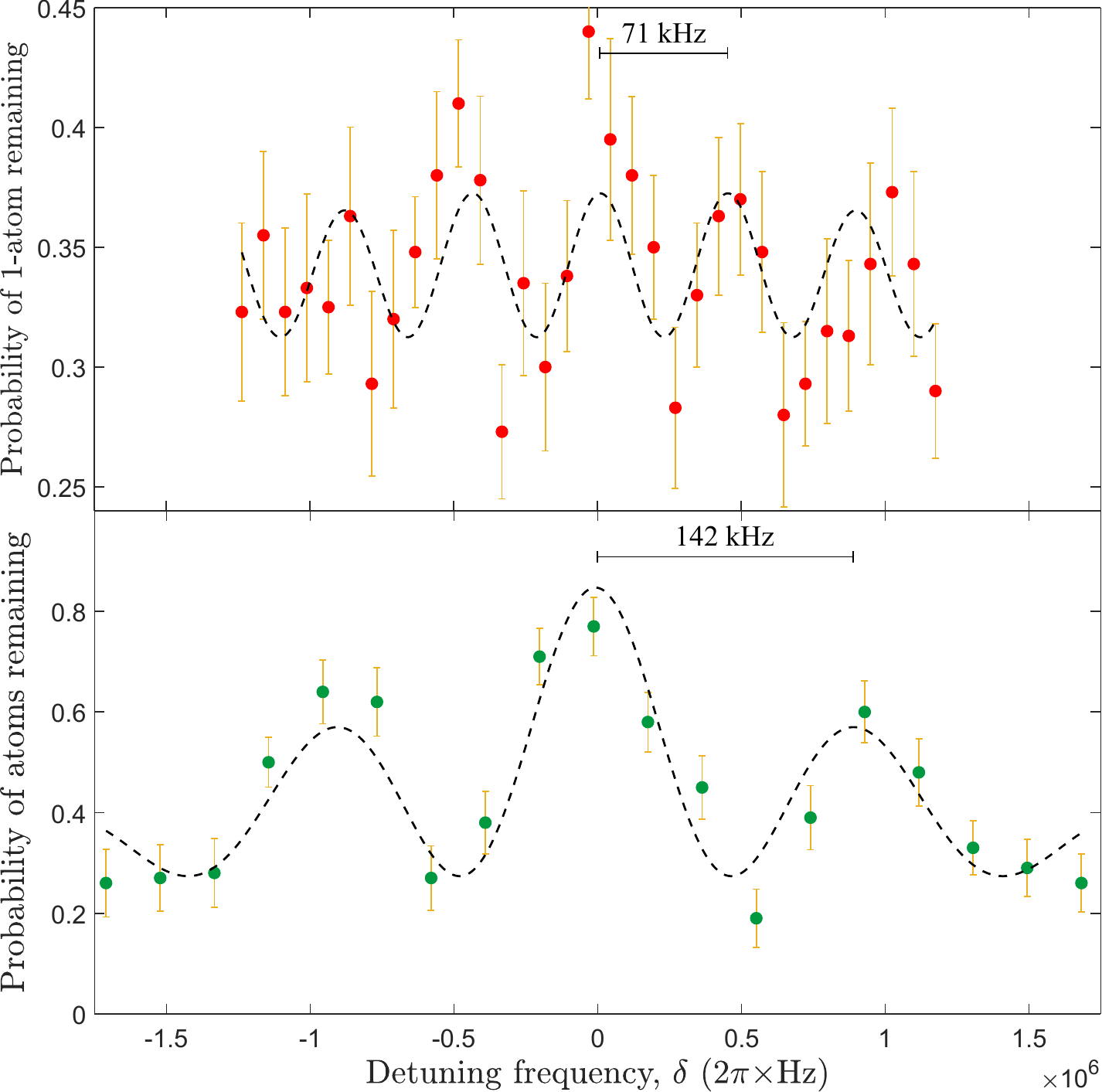}
\caption{\label{fig:Ramsey Entangled state}
Results of the $\pi/2-T-\pi/2$ Ramsey-type experiment: The top panel presents Ramsey fringe obtained with the entangled state%, showing the probability of 1-atom remaining in the trap as a function of Raman detuning $\delta$. 
Dashed line: The expected signal for two entangled atoms (Eq.~17 in \cite{supplementary}) with a fitted offset and fringe contrast. A $\chi^2$-test accept the model. Bottom panel: The result from two unentangled atoms. The dashed line is Eq.~18 in \cite{supplementary} with fitted offset and fringe contrast. }
\end{figure}
Figure~\ref{fig:Ramsey Entangled state} (top panel) shows the probability of detecting one atom remaining in the trap as a function of the Raman detuning $\delta$ for a wait time of \SI{5}{\micro\second} between the $\pi/2$ pulses. It displays the characteristic Ramsey fringes. The frequency difference between the fringes is \SI{71}{\kilo\hertz}. 
The bottom panel shows the result of the Ramsey fringes obtained by using unentangled atoms, both initially prepared in $|1\rangle$. The spacing between the fringes is twice that found with the entangled state. The narrower Ramsey fringes indicate that the entangled state can give a factor of two improvement in the sensitivity of measurements of the Zeeman splitting, or, in other words, of the magnetic field.
The reason for the factor of two improvement stems from the fact that the spin-state after the first $\pi/2$ pulse ($-\frac{i}{\sqrt{2}}\left( \left|1, 1 \right\rangle + \left|-1, -1 \right\rangle \right)$) has twice the Zeeman energy difference of a single atom in a superposition of the $\left|1 \right\rangle$ and $\left|-1 \right\rangle$ states, and acquires twice the relative phase in the wait period. 
We stress that while Fig.~\ref{fig:Ramsey Entangled state} provides a proof-of-principle that the generated entanglement is useful, it does not provide an entanglement-enhanced measurement in itself. A main reason for that is the lower \mfa{probability (about 0.37)} with which the entangled state is prepared compared to the initial state for the unentangled pair. In principle, this can be overcome by ejecting atoms in $\left| 0 \right\rangle$ and $\left| \pm 2 \right\rangle$, thereby removing runs that did not generate the desired state. 

\mfa{\textit{Discussion.---} Several methods are able to entangle the internal states of laser cooled atoms or ions, despite these having residual thermal motion. Most prominently the M{\o}lmer-S{\o}rensen gate and the Rydberg gate, which power ion- and neutral-atom-based quantum computing efforts. These use laser driven coherent dynamics, that is insensitive to the motional state, to transfer atoms or ions to an entangled state. Their high fidelity is impressive, but the simplicity and robustness of using a thermal relaxation process may also be useful.
This is not limited to efforts to use arrays of individual atoms in optical tweezers for magnetic field imaging \cite{Schaffner2024}, but can be considered in a wider context.
}

\textit{Conclusions and outlook.---}  We experimentally demonstrated that coupling a quantum system to a hot environment can drive the system from an initially unentangled state into an entangled state. Explicitly, we showed that the coupling between the spin-state of two atoms in an optical tweezer and their motion (provided by spin-changing collisions) can drive the atom pair into an entangled spin-state \mfa{even when the motion is in a thermal state}. Moreover, we showed that the spin-entanglement generated is metrologically useful as it can enhance magnetic field measurements.
In future, it could be interesting to investigate whether the entanglement is sufficiently robust to survive subsequent separation of the two atoms into physically-separated optical tweezers.
A deeper understanding of the spin-dynamics may also reveal if schemes that prepare the entangled state with higher %fidelity 
\mfa{probability }are possible. 

\textit{Acknowledgments.---} 
This work was supported in part by the Marsden Fund Council from Government funding, administered by the Royal Society of New Zealand (Contract No. UOO1835). Additionally, it was supported by Quantum Technologies Aotearoa, a research programme of Te Whai Ao – the Dodd Walls Centre, funded by the New Zealand Ministry of Business Innovation and Employment through International Science Partnerships, contract number UOO2347. SSS was supported by an Australian Research Council Discovery Early Career Researcher Award (DECRA), Project No. DE200100495. The contribution of SP was supported in part by grant NSF PHY-2309135 to the Kavli Institute for Theoretical Physics (KITP).

\appendix*

\section{Experimental details}
We load two $^{85}$Rb atoms into separate optical tweezers using the procedure described in Refs.~\cite{carpentier2013,fung2015, grunzweig2010}. The tweezers are two linearly polarized \SI{1064}{\nano\meter} laser beams focused to a spot size of $\omega_0=\SI{1}{\micro\meter}$ separated by \SI{4}{\micro\meter}. At this stage the tweezer depth is $h\times\SI{57}{\mega\hertz}$. We then image the atoms to confirm their presence \cite{sompet2019,hilliard2015}. Approximately 60\% of the attempts result in successfully loading two atoms (one atom in each optical tweezer); failed attempts are excluded from consideration. Next, polarization gradient cooling lowers the atoms' temperature, and optical pumping followed by a Raman transfer prepares them in the $F=2, m_F=0$ groundstate, achieving an efficiency of 93\% \cite{sompet2019}. Throughout the optical pumping and Raman transfer, a bias magnetic field of \SI{8.5}{\gauss} defines the quantization axis and controls the spin dynamics of the atom pair.

We near-adiabatically merge the atoms by moving one tweezer closer to the other using an acousto-optic modulator to control the position and power of the tweezer beams over a \SI{20}{\milli\second} duration \cite{Andersen2022}. Following the merge, we adiabatically ramp off the power of the moved tweezer beam during \SI{20}{\milli\second}. At the same time, we ramp the other optical tweezer to the desired trap depth and set the bias magnetic field to the specified value of \SI{0.5}{\gauss}. This procedure results in both atoms residing in a $h\times$\SI{22}{\mega\hertz} deep optical tweezer, with a temperature of $\sim$\SI{40}{\micro\kelvin} (measured using the release-recapture technique). 
Spin-changing collisions between the atoms now generate the $m_F$ population dynamics illustrated in Fig.~\ref{fig:Process} (b).

 To eject atoms in $\left| 1 \right\rangle$ we selectively transfer them to the $|F=3,m_F=1\rangle$ state using a Raman transition and ejecting atoms into the $F=3$ state using the push-out technique.  

\nocite{*}

\bibliography{References_Main}% Produces the bibliography via BibTeX.

\end{document}